\documentclass[galaxies,review,accept,oneauthor,pdftex]{Definitions/mdpi} 

\firstpage{1} 
\pubvolume{1}
\issuenum{1}
\articlenumber{0}
\pubyear{2021}
\copyrightyear{2021}
\externaleditor{Academic Editor: Margo Aller, Jose L. Gómez and Eric Perlman}
\datereceived{8 April 2021} 
\dateaccepted{12 May 2021} 
\datepublished{} 
\hreflink{https://doi.org/} 

\newcommand\T{\rule{0pt}{2.6ex}}       
\newcommand\B{\rule[-1.2ex]{0pt}{0pt}} 

\Title{Probing Magnetic Fields and Acceleration Mechanisms in Blazar Jets with X-ray Polarimetry}

\TitleCitation{Probing Magnetic Fields and Acceleration Mechanisms in Blazar Jets with X-ray Polarimetry}

\Author{Fabrizio Tavecchio \orcidA{}}

\AuthorNames{{Fabrizio Tavecchio}}

\AuthorCitation{Tavecchio, F.}

\address[1]{INAF -- Osservatorio Astronomico di Brera, Via E. Bianchi 46, I-23807 Merate (LC), Italy; fabrizio.tavecchio@inaf.it 
}

\abstract{X-ray polarimetry promises us an unprecedented look at the structure of magnetic fields and on the processes at the base of acceleration of particles up to ultrarelativistic energies in relativistic jets. Crucial pieces of information are expected from observations of blazars (that are characterized by the presence of a jet pointing close to the Earth), in particular of the subclass defined by a  synchrotron emission extending to the X-ray band (so-called high synchrotron peak blazars, HSP). In this review, I give an account of some of the models and numerical simulations developed to predict the polarimetric properties of HSP at high energy, contrasting the predictions of scenarios assuming  particle acceleration  at shock fronts with those that are based on magnetic reconnection, and I discuss the prospects for the observations of the upcoming  Imaging X-ray Polarimetry Explorer ({\it IXPE}) satellite.}


\begin{document}
\section{Introduction}

The comprehension of the physical processes behind  the acceleration of relativistic particles (electrons, possibly nuclei) in the relativistic collimated outflows (jets) that are associated with Active Galactic Nuclei (AGN) is one of the most intriguing challenges of modern astrophysics. In fact, the energization of charged particles at (ultra)relativistic energies involves strictly interlaced processes played by magnetic fields, plasma and related instabilities (e.g.,  \cite{bykov12, romero17, bell20}). Moreover
, relativistic jets are structures of enormous complexity, where the interplay of several processes, acting at spatial and temporal scales spanning several orders of magnitude, produces a rich phenomenology, as witnessed by observations (e.g., \cite{blandford19}). While particle acceleration occurs all along the jet, from subpc to the Mpc scale, remarkable effort has been dedicated to blazars, whose observed emission is likely produced within the first parsecs from the black hole \cite{rani18}. In these sources, the jet is almost pointed toward the Earth and the consequent relativistic beaming causes a strong amplification of the non-thermal emission, which covers the whole electromagnetic spectrum (e.g., \cite{romero17}).

Current studies converge to support two main general paths by which the energy that is carried by a relativistic jet can be dissipated and made available for the acceleration of relativistic particles: {\it (a)} for {\it magnetically dominated} jets (i.e., jets with a magnetization parameter $\sigma>1$, where we define $\sigma=B^2/4\pi \rho c^2$, with $B$ being the magnetic field and $\rho $ as the plasma density) simulations show that a sizable part of the initial magnetic energy can be dissipated through (relativistic) reconnection, being easily triggered during the non-linear stages of jet instabilities (e.g., Kelvin--Helmholtz or current-driven kink instabilities, (e.g., \cite{hardee13})). Particle-in-cell (PIC) simulations show that, in current sheets associated with magnetic reconnection, particles can be efficiently accelerated forming non-thermal energy distributions (e.g., \cite{zenitani01, sironi14, guo15, werner16}); {\it (b)} for {\it weakly  magnetized} flows, instead, the most likely dissipation sites are shocks (e.g., \cite{marscher85, aller1985,baring17}), where the formation of non-thermal populations  occurs through the classical diffusive shock acceleration (DSA) mechanism (e.g., \cite{blandford87, bell14, spitkovsky08}).

Clearly, while simulations and theoretical studies can delineate the landscape of the potential physical processes at action, only the confrontation with the observational evidence can decide which mechanism(s) is ultimately responsible for particle acceleration in jets. In this respect, for a long time, multiband polarimetric measurements have been considered to be a powerful tool for investigating  structure and dynamics of relativistic jets, magnetic field geometries and particle acceleration (e.g., \cite{angel80, blandford19}). The regular multiwavelength monitoring of blazars, which was greatly intensified in the last decade after the advent of {\it Fermi}-LAT, led to the identification of regular or common variability patterns, often involving polarimetric properties. The evidence for systematic and large ($\sim$180 degrees) variations of the polarization angle appears to be particularly important,  being often associated with gamma-ray flares (e.g., \cite{blinov15, blinov18}). Possible interpretations advanced to explain these observations include an emission region moving along a helical path in a jet dominated by a toroidal field (e.g., \cite{marscher08, marscher10, larionov13}, a jet bending at parsec scales \cite{abdo10, nalewaiko10}, turbulence in the flow (usually described in terms of stochastic models, (e.g., \cite{kiehlmann17}), possibly generated downstream of a standing \mbox{shock \cite{marscher14,marscher15}}.


In this review, I intend to focus on the potentialities offered to the study of the acceleration mechanisms in jets by the new-generation of X-ray polarimeters, in particular the forthcoming Imaging X-ray Polarimetry Explorer ({\it IXPE}) satellite \cite{weisskopf16}. Specifically, we will focus our attention to the subclass of blazars characterized by the synchrotron component peaking in the X-ray band, the so-called high synchrotron peak (HSP). Polarization measurements of these sources in the X-rays, exploring the most energetic, freshly accelerated, electrons, can provide unique {\it in situ} information on magnetic field geometry, turbulence, and particle distribution inside the jets, key inputs to test and improve our models. 

After a brief introduction of the two potential acceleration processes (magnetic reconnection and DSA), I will discuss some recent studies and simulations that were devoted to the identification of the polarimetric signatures expected in the two scenarios and the perspective to test and constrain these models by observations with the upcoming Imaging X-ray Polarimetry Explorer ({\it IXPE}) satellite.

\section{High Synchrotron Peak Blazars as Laboratories for Particle Acceleration}

Before concentrating on the potentialities of polarimetric studies of blazars in the X-ray band, it is worth illustrating the motivations to focus our attention on the particular class of HSP. 

\begin{figure}[H]

 \includegraphics[width=10.5 cm]{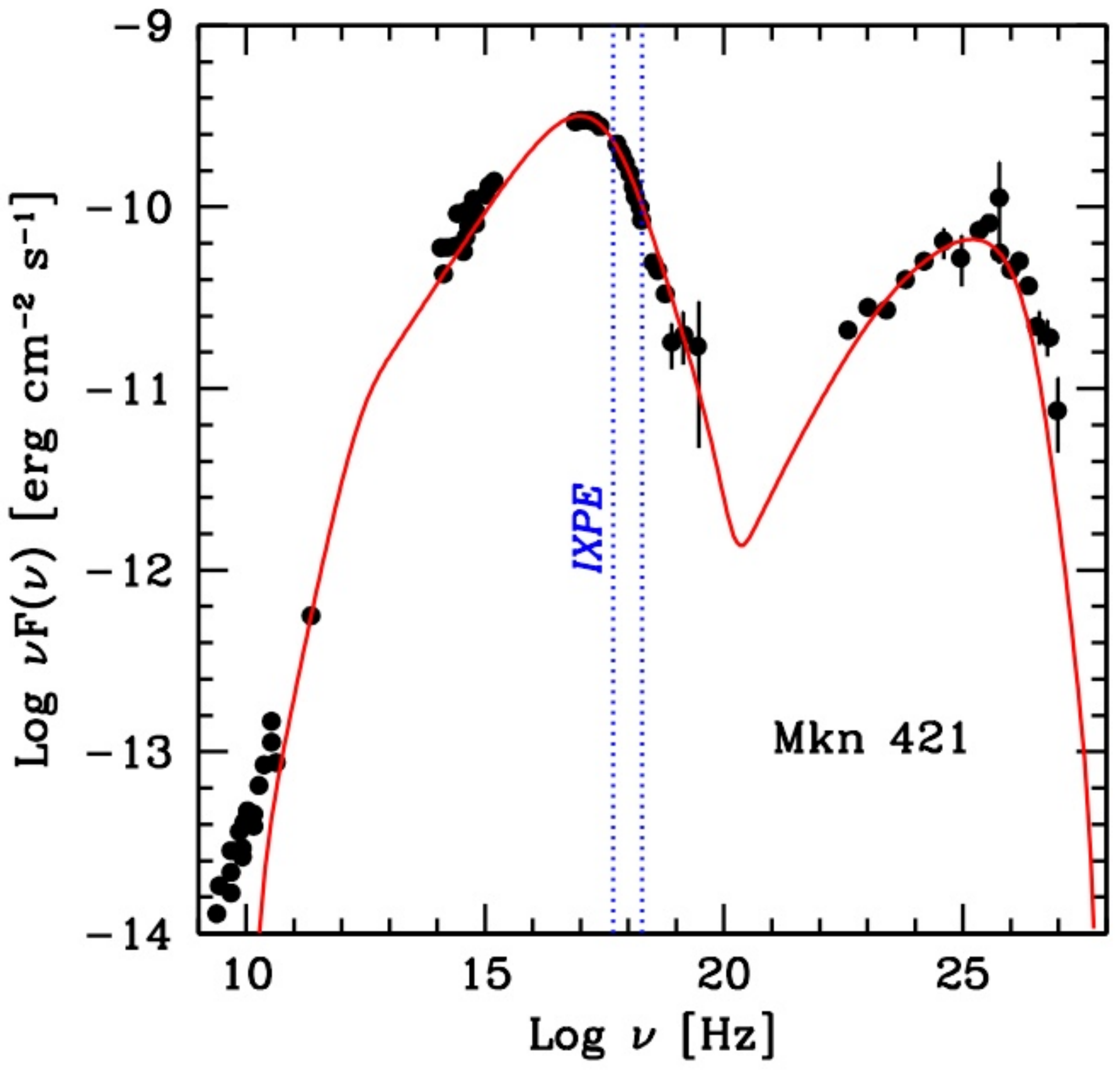}

\caption{The broadband spectral energy distribution of the prototypical HSP blazar Mkn 421 in a low-medium state (data from \cite{abdo11}). The red line represents the emission calculated with a one-zone synchrotron-self Compton model \cite{tavecchio16}. Vertical blue dotted lines define the energy range covered by the X-ray polarimeter of the upcoming {\it IXPE} satellite, corresponding to the synchrotron emission from the most energetic electrons.  
}  
\label{fig:mkn421}
\end{figure}

Figure~\ref {fig:mkn421} reports the spectral energy distribution (SED) of the prototypical HSP Mkn 421. The two bumps, interpreted as produced through synchrotron and inverse Compton emission (for the hadronic interpretation of the high-energy component, see, e.g., \cite{reimer03}) peak around 1 keV and 100 GeV, respectively. The high-energy component displays a tail extending up to several TeV. Indeed, HSP are the most abundant sources that are detected at TeV energies by the current Cherenkov telescopes.

In the simplest emission scenarios, the observed emission of HSP is reproduced by assuming a single, compact emission region, which is homogeneously filled by a non-thermal population of relativistic electrons and tangled magnetic field. Because of the paucity of nuclear radiation fields in the core of HSP, it is largely assumed that the inverse Compton scattering occurs on the synchrotron photons. In this greatly simplified model (one zone SSC, synchrotron self-Compton), the degrees of freedom of the system are quite limited and the observed SED allows us to greatly constrain the physical properties of the emission region (e.g., \cite{tavecchio98, tavecchio10}). In particular, relatively low magnetic fields ($B\lesssim 0.1$ G) are generally derived (often well below equipartition with the radiating electrons, \cite{tavecchio16}), implying that the electrons emitting synchrotron radiation in the X-ray band ($h\nu_X=1$--$10$ keV) are characterized by large Lorentz factors, of the order of $\gamma_{\rm X}=(2\pi m_ec\nu_X/eB\delta)^{1/2} \sim 10^5$--$10^6$, where $\delta\approx 10$ is the relativistic Doppler factor. These large energies determine severe radiative losses and, thus, very short radiative cooling times for these electrons ($t_{\rm cool}=7.8\times 10^5 B_{-1}^{-2} \, \gamma _{5}^{-1}$ s). Because of the short lifetime, during the emission electrons can propagate for very small distances, which implies that the synchrotron X-ray radiation must be produced very close to the region where particles gain their energy. This important conclusion explains why the study of the polarimetric properties of the X-ray emission of HSP can provide invaluable information on the acceleration processes. Another favorable property of HSP is that the X-ray emission, close to the peak of the synchrotron hump, is very bright (often exceeding $10^{-10}$ erg cm$^{-2}$ s$^{-1}$ in Mkn 421, see Figure~\ref{fig:mkn421}), thus allowing for precise polarimetric measurements, even with relatively short exposures ($\sim$10$^{3}$ s, e.g., \cite{tavecchio20}). 

\section{Particle Acceleration in Blazar Jets: Shocks and Magnetic Reconnection}

Acceleration at shock fronts has been classically considered to be the prime candidate mechanism behind particle energization in several astrophysical environments---from the solar system to cluster of galaxies \cite{blandford87,bell14}. For jets, in particular those that are associated with blazars, early works already identified shocks (travelling or standing) as the sites where the energetic emission is produced (e.g., \cite{marscher78, blandford79}). Indeed, shocks are predicted to form quite naturally in the supersonic flows characterizing relativistic jets. In particular, so-called internal shocks are easily produced in unsteady flows, while (oblique) reconfinement shocks mark the position where the external medium drives the recollimation of an underpressured expanding jet. In a real jet, both kinds of shocks are likely to occur. Models of blazar emission are usually built on the assumption of the existence of a generic emission zone (often implicitly identified with the downstream shock region) filled by a non-thermal population of particles whose physical origin is usually not defined (e.g., \cite{boettcher07,tavecchio10}). More refined models attempting to link the properties of the observed emission to a more realistic physical picture have been recently developed and used to infer the physical characteristics of the shock  (e.g., \cite{baring17, boettcher19}). 

A point worth mentioning here concerns the role of turbulence. It is well known that a sufficiently high level of disorder of the magnetic field lines (allowing for the rapid diffusion of particles in the downstream flow to counterbalance advection) is generally required for the DSA to work \cite{bell14}. In fact, turbulence is thought to be induced by the accelerated particles themselves, which are able to excite Alfven waves in the flow (e.g., \cite{schure12}). Studies that are based on the modeling of blazar emission suggest that turbulence could decay quite rapidly after the shock \cite{baring17}. As we will discuss below, turbulence, disturbing the order of the magnetic field, is an important player in polarimetric studies, with a strong impact on the expected degree of polarization of the emitted radiation.

Recent investigations \cite{sironi15} pointed out that, for jets with sufficiently high magnetization ($\sigma>0.1$), DSA is relatively inefficient and can only produce a significant non-thermal component for mildly relativistic shocks and under special configurations in which the magnetic field lines in the upstream flow are nearly orthogonal to the shock front (i.e., parallel
shocks). This, together with other lines of evidence, concur to support the view that, besides shocks, current sheets associated with magnetic reconnection sites can potentially play an important role in accelerating the particles that produce the strong non-thermal emission from jets. In fact, (1) MHD simulations agree on the fact that jets start as magnetically dominated outflows in which the magnetic energy is progressively converted to kinetic energy while the jet accelerates (e.g., \cite{komissarov07,tchek09}). As it is quite well known, (2) magnetically dominated jets are prone to several kinds of instabilities, in particular the current-driven kink instability whose non-linear stages create the conditions for efficient dissipation of magnetic energy through magnetic reconnection. (3) PIC simulations convincingly show that current sheets formed in highly magnetized plasmas are sites of fast, relativistic reconnection that can sustain efficient acceleration of particles with (possibly anisotropic, \cite{zhdankin20}) non-thermal distributions (e.g., \cite{zenitani01, sironi14, guo15, werner16}). An alternative to reconnection induced by instabilities is a scenario assuming a striped-wind jet, with a magnetic field organized in layers of opposite polarity 
\cite{giannios19,zhang21}. The continuous dissipation of magnetic field through reconnection at the layer interfaces naturally drives the acceleration of the jet and particle acceleration.

\begin{figure}[H]

    \includegraphics[width=14. cm]{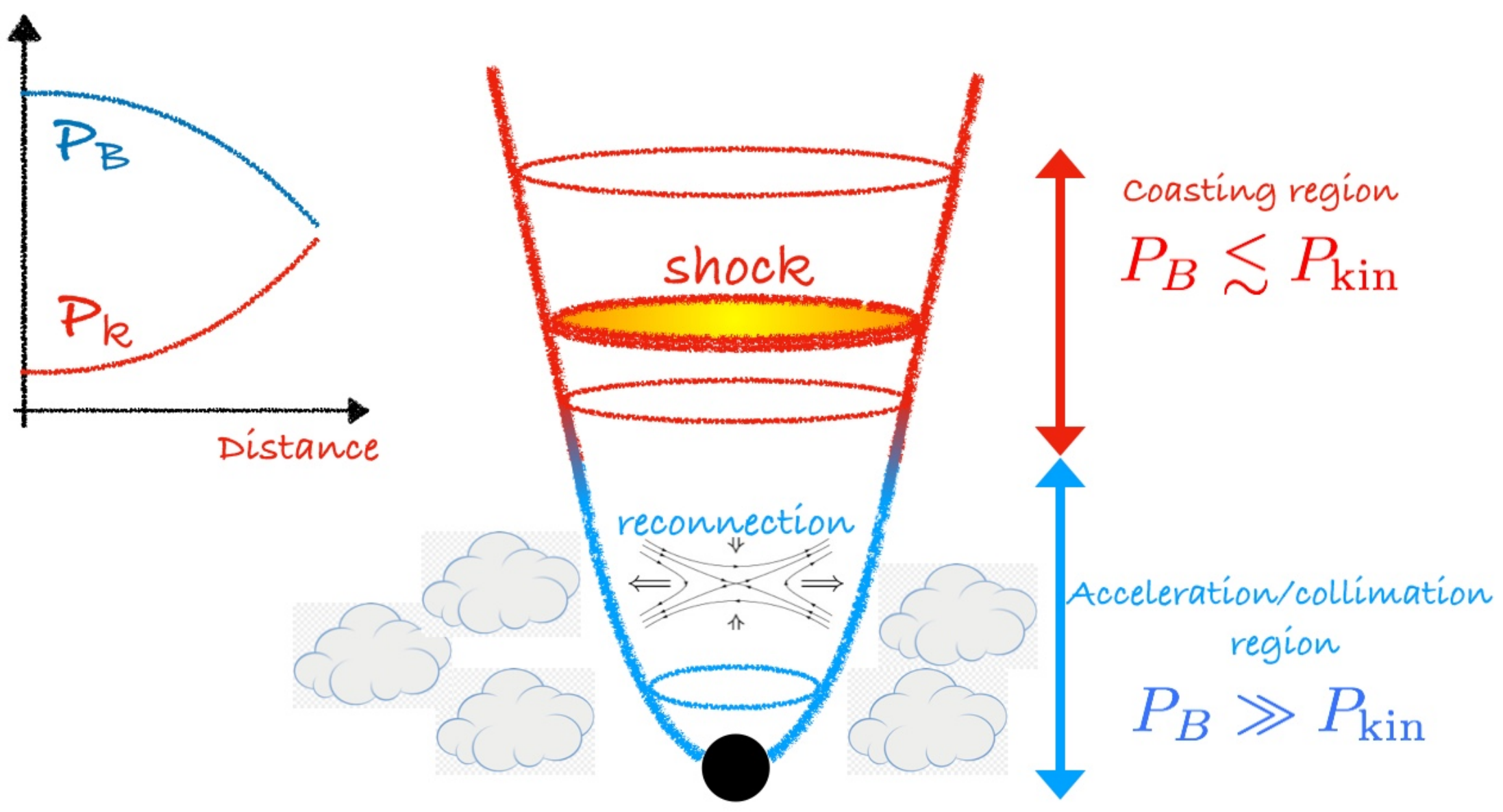}

\caption{A possible global scenario for dissipation and particle acceleration in relativistic jets. The jet starts as a magnetically dominated flow which progressively accelerates. In this region the main channel for energy dissipation is magnetic reconnection, likely triggered by current-driven instabilities. During the acceleration phase, the magnetic energy is converted into kinetic bulk energy (schematic plot on the left) until substantial equipartition is reached.  At larger distances the reduced magnetization of the flow allows efficient particle acceleration by shocks.}  
\label{fig:jetprofile}
\end{figure}

It is important to remark that the two mechanisms mentioned above (i.e., shock and magnetic reconnection) are not mutually exclusive and they can simultaneously operate in the same jet, possibly at different distances (i.e., reconnection in the inner, magnetically dominated regions, DSA at large scales, where the jet is likely matter dominated, (e.g., \cite{bell14})). Clearly, this would increase the degree of complexity of the system (Figure~\ref{fig:jetprofile}).

\section{Modeling X-ray Polarimetric Signatures in HSP}

Theoretical studies and numerical simulations are increasing and extending our understanding of the intricate network of physical processes behind the acceleration of relativistic particles in jets. However, despite the important efforts, the gained knowledge is still too limited to make precise predictions to be compared with observations. Perhaps the most important difficulty is represented by the huge range of spatial and temporal scales that are involved. In fact, while PIC simulations offer a detailed description of the dynamics at very small scales (including particle acceleration), the emission from jets involves their global structure and dynamics (i.e., instabilities, shocks), which are best captured by a fluid MHD treatment. 

In order to produce an approximate view of the expected emission, it is therefore useful to adopt simplified models that can incorporate some of the results obtained through simulations, but that are also suitable to a simple analysis. In the following, I will try to offer a glimpse of the current research along these lines and of some of the most relevant results for polarimetry in the X-ray band.

\subsection{Shocks: Large Scale vs. Small Scale Treatments}

The polarimetric signatures that are associated with shocks in jets have been explored in several works (e.g., \cite{laing80, hughes85,jones88,cawthorne90,nalewajko12,zhang16}). These studies are generally based on the MHD description and consider the macroscopic properties of plasma and magnetic field in the downstream region to characterize the emitted radiation. A general feature is that, due to the compression that is operated by the shock, the field downstream of the shock develops a predominant component parallel to the shock front, regulating the polarization of the emitted radiation. 

Recent studies demonstrated that models based on MHD shocks are able to  reproduce  part of the observed phenomenology, for instance the large polarization angle swings that are often observed in powerful blazars (possibly caused by light travel time effects within an underlying axisymmetric emission region \cite{zhang15,zhang16}), although they cannot explain angle rotations of more than 180 degree or rotations with different directions in the same source. Dedicated MHD simulations (assuming helical force-free magnetic fields) reveal that the agreement with the phenomenology that is displayed by the optical polarization is possible only for highly-magnetized jets \cite{zhang16}. In fact, for low magnetization ($\sigma\lesssim 0.1$) the resulting degree of polarization during the late phases of a flare reaches quite high values ($\Pi\simeq 40$--$50\%$), in contrast with the usual observed behavior. A better agreement with observations (i.e., a degree of polarization approaching a few percents during decay phases) is reached when large magnetization ($\sigma$$\sim$10) is considered. This result appears in remarkable contrast with the small acceleration efficiency anticipated for shocks in highly magnetized flows \cite{sironi15}.

Models that are based on a large-scale approach are surely valid for the low-energy (e.g., optical in HSP blazars) emission, produced by long-living electrons that can travel to large distances after being accelerated and, therefore, experience an average field that is well represented by the MHD treatment. However, this approach neglects two ingredients that, likely not greatly affecting the properties of the optical emission, can critically shape the emission at higher energies, in particular in the X-ray band. In fact, (1) the severe cooling limits the propagation of the most energetic electrons to a very thin layer after the shock front. Therefore (2), the complex small-scale structure of the magnetic field close to the shock, not captured by large-scale MHD simulations, has to be properly considered in deriving the polarimetric properties.

\begin{figure}[H]
    \includegraphics[width=7.2 cm]{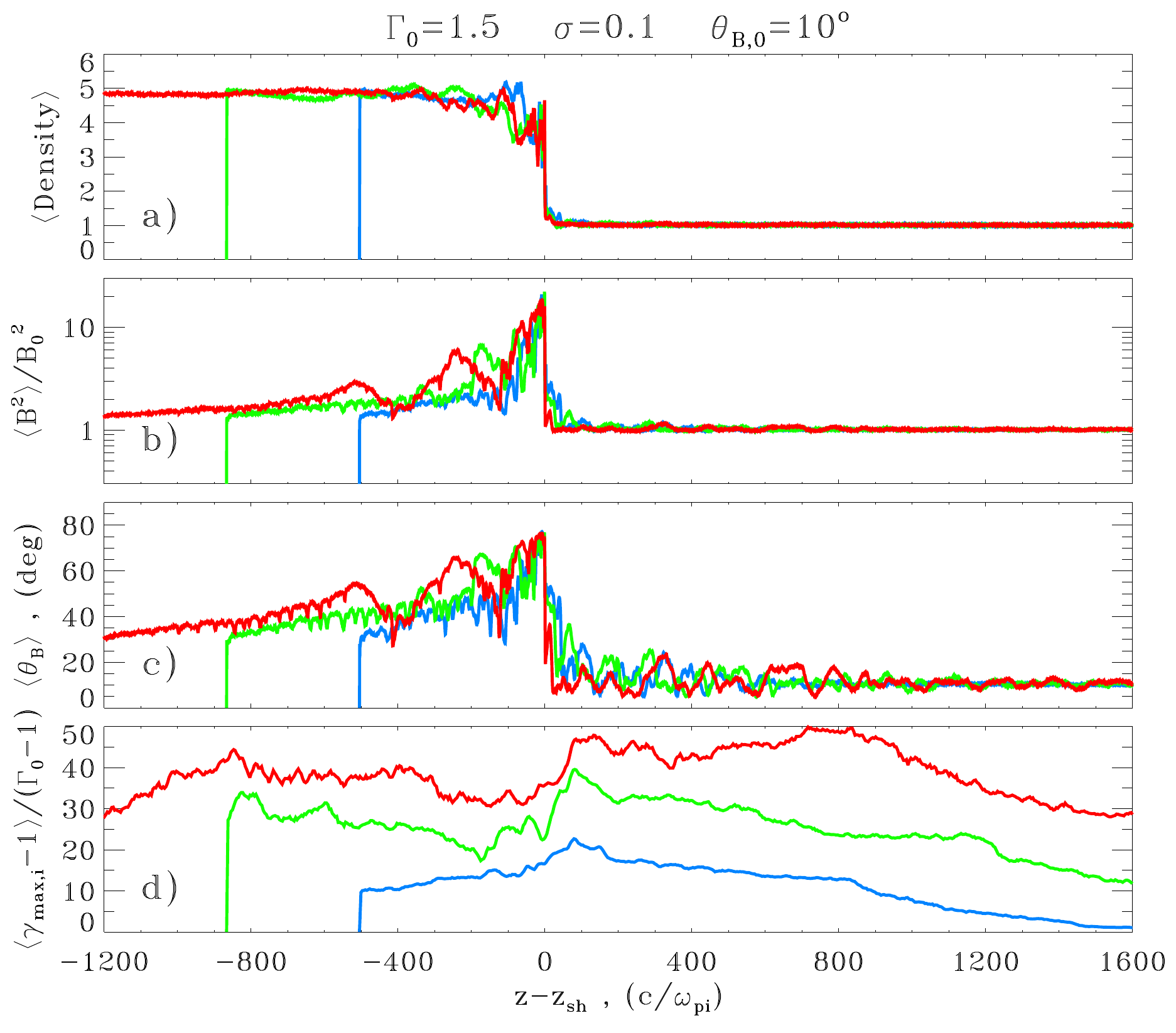}\\
\includegraphics[width=10.3 cm]{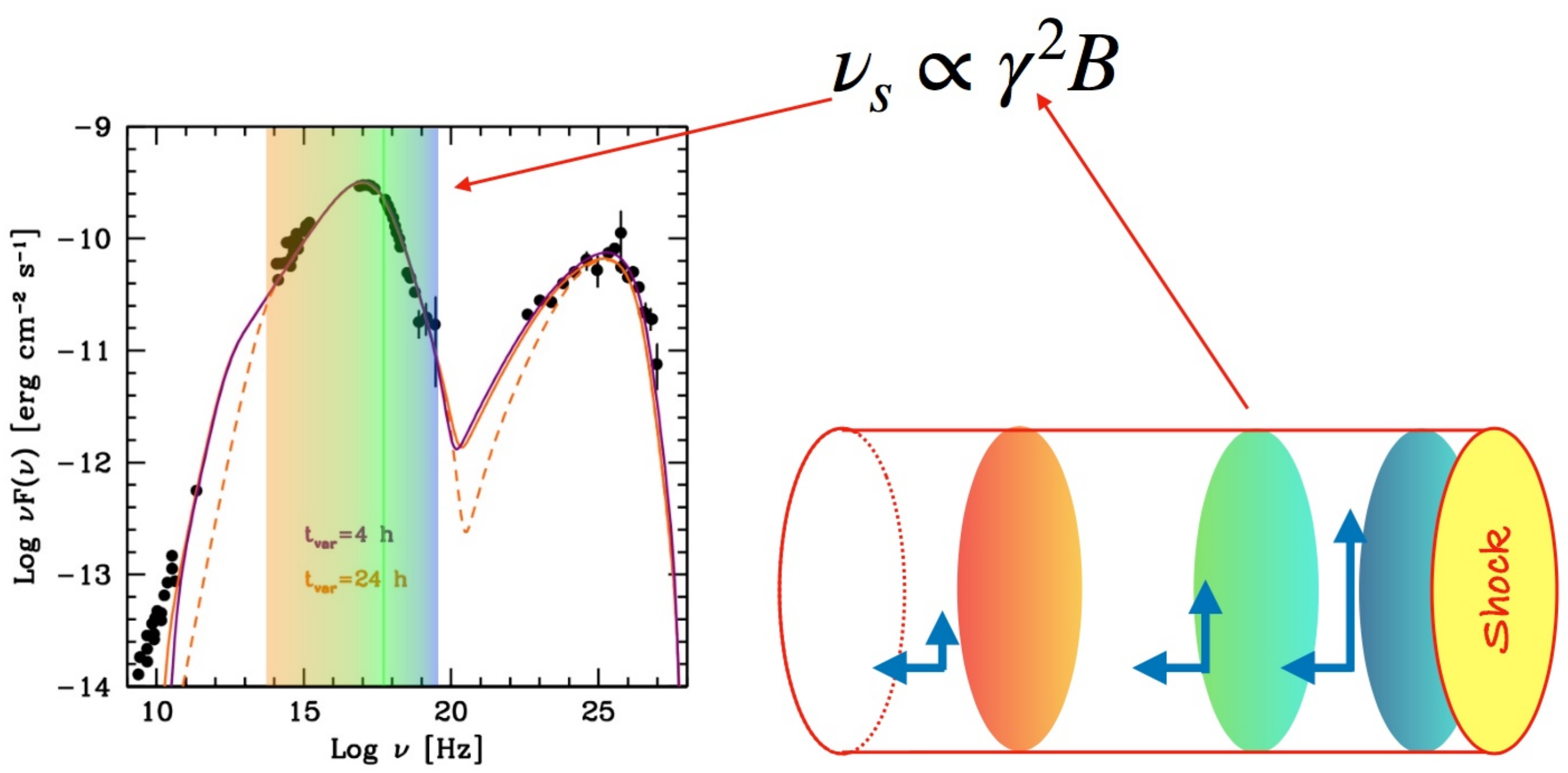}
\caption{(\textbf{Upper panel:}) The results of a PIC simulations of a parallel, mildly relativistic, low magnetization shock. The panel reports (top to bottom) the spatial profile of density, magnetic energy normalized to that carried by the upstream flow,  average angle between the magnetic field and the shock normal and  average Lorentz factor of the accelerated electrons. The second panel clearly shows the generation of a strong magnetic field at the shock parallel to the front (third panel). From \cite{tavecchio18}. Reproduced with kind permission from Oxford University Press and the Royal Astronomical Society. (\textbf{Lower panel}) Cartoon of the model, bei g inspired by PIC simulations, assumed to calculate the expected polarization for the shock scenario. Electrons that are accelerated at the shock (right) are advected downstream and cool through synchrotron and SSC emission. Electrons emitting in the X-ray band (blue), which have a quite limited lifetime, emit very close to the shock front, where the magnetic field is dominated by the self-generated component parallel to the shock (blue arrows). Being produced in a region with a well defined direction of the field, the synchrotron emission from these electrons is highly polarized.
Electrons with lower energy (green and orange) produce synchrotron radiation at a lower frequency (UV, optical) and, because of the longer cooling timescale, can propagate at distances where the self-generated field decays below the field carried by the flow, which are assumed to be nearly perpendicular to the shock. The integrated emission from these low energy electrons originates in a region in which the two (orthogonal) components of the field are, on average, comparable, which leads to a low degree of~polarization.
\label{fig:shock}}
\end{figure}

In fact, PIC simulations of shocks confirm early predictions (e.g., \cite{schure12}) that streaming suprathermal ions excite circularly polarized Alfven waves that, in the downstream compressed flow, sufficiently close to the front, emerge as an intense magnetic field component parallel to the shock front (e.g., \cite{caprioli14}). As an example, we report (see Figure~\ref{fig:shock}, upper panel) the results of a simulations starting with a weakly magnetized ($\sigma=0.1$) upstream flow with a magnetic field that is slightly inclined with an angle of 10 deg with the shock normal. The second panel shows the magnetic energy (normalized to the value in upstream flow). The presence of the self-produced field that is close to the shock is rather evident, as well as (third panel) the fact that the field is prevalently parallel to the shock (or orthogonal to the axis). At the shock, the self-generated field contains an energy approximately 20 times that of the original (almost perpendicular) field. However, this self-produced field component decays relatively quickly with distance, and sufficiently far from the shock, the magnetic field recovers the MHD expectations.

\begin{figure}[H]

    \includegraphics[width=10. cm]{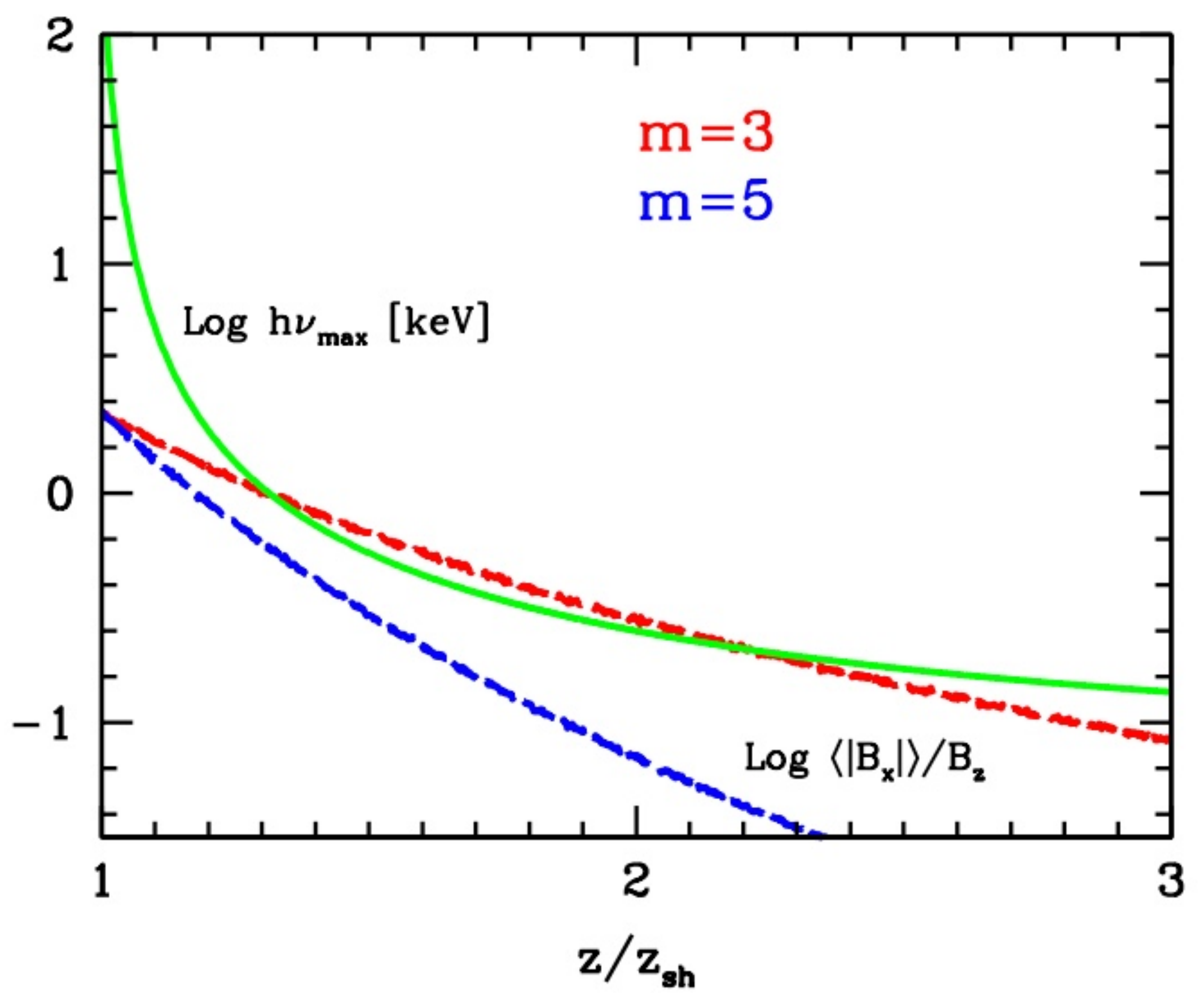}

\caption{The spatial profiles of the ratio between the averaged perpendicular component of the magnetic field projected onto the plane of the sky and the parallel magnetic field component as a function of the distance from the shock front for the model that si assumed in \cite{tavecchio18,tavecchio20}. The profiles correspond to $m=3$ (red) and $m=5$ (blue). The solid curve shows the maximum synchrotron photon energy (in the observer frame) as a function of the distance from the shock front. The synchrotron radiation in the X-ray range ($h\nu_{\rm max} > 1$ keV) is  produced in the region where the field is dominated by the component parallel to the shock front, causing a large degree of polarization. From \cite{tavecchio20}. Reproduced with kind permission from Oxford University Press and the Royal Astronomical Society.}  
\label{fig:profile}
\end{figure}

Being informed by these results, References \cite{tavecchio18,tavecchio20} presented a model for the polarization of the synchrotron radiation produced by high-energy particles accelerated at a shock---being particularly suitable for describing the expected properties of the X-ray emission. The adopted  set-up is simple: the jet is modeled as a cylinder, and it is assumed that the observer lies at an angle $\theta_{\rm v}=1/\Gamma_{\rm d}$, where $\Gamma_{\rm d}$ is the bulk Lorentz factor of the downstream plasma. In this condition, in the downstream frame the line of sight is perpendicular to the jet axis. A key ingredient taken into account in the model is the different cooling length that is associated with electrons of different energies (and, therefore, emitting at different frequencies). Indeed, since the cooling time of an electron of energy $E$ scales as $t_{\rm cool}\propto 1/E$, electrons emitting, for instance, in the X-rays have a quite smaller lifetime than those producing optical radiation. Because electrons are assumed to be advected by the downstream flow with speed $v_{\rm adv}$, the emission of the X-ray radiation is concentrated in a very thin layer of thickness $\lambda _{\rm cool} \simeq v_{\rm adv}t_{\rm cool}$ after the shock front, while, because of the longer lifetime of the corresponding electrons, the optical radiation is produced in a much larger volume (see the cartoon in Figure~\ref{fig:shock}, ower panel). In Figure~\ref{fig:profile}, I report an example of the profile of the maximum synchrotron frequency in the downstream region (see \cite{tavecchio20} for details on the adopted parameters). Clearly, X-rays ($E>1$ keV) are only produced at a small distance from the shock.

The energy-dependent stratification of the electrons, together with the presence of the strong, decaying self-generated field, shape the polarimetric properties of the emitted synchrotron radiation. The  radiation that is produced by particles at high energy (i.e., in the X-ray band), close to the front---where the projected magnetic field is well defined, being largely dominated by the parallel self-generated component (see the blue and red lines in Figure~\ref{fig:profile})---will acquire a large degree of polarization. On the other hand, particles of low energy, while advected by the downstream flow, shine over a large portion of the downstream volume. Because the self-generated field only dominates in the close vicinity of the shock, the low-energy (e.g., optical) emission that is produced by these electrons will become less polarized (see also \cite{angelakis16, itoh16}, which, however, do not consider the role of the self-generated fields).

\begin{figure}[H]
\widefigure

\includegraphics[width=7.5 cm]{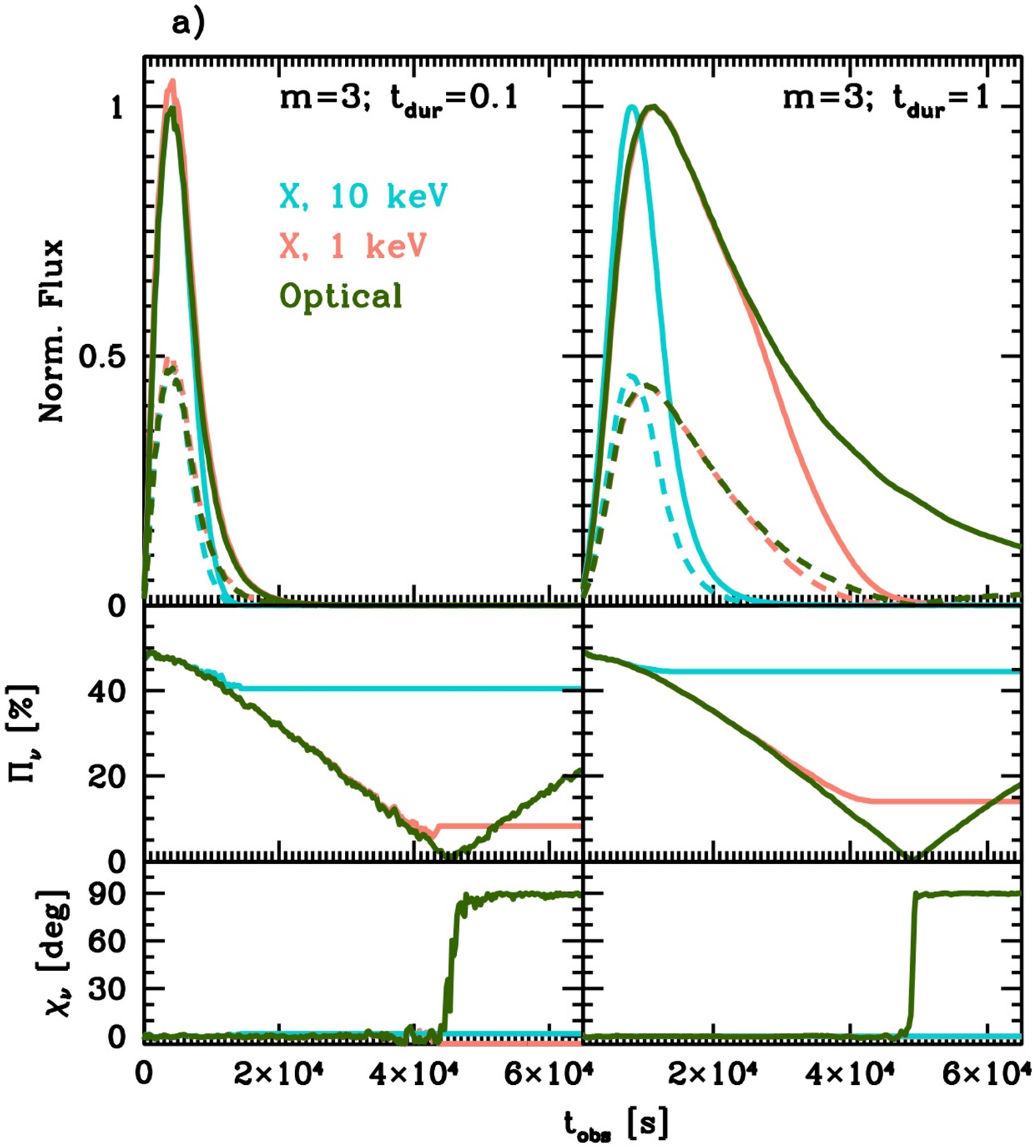}
\includegraphics[width=7.5 cm]{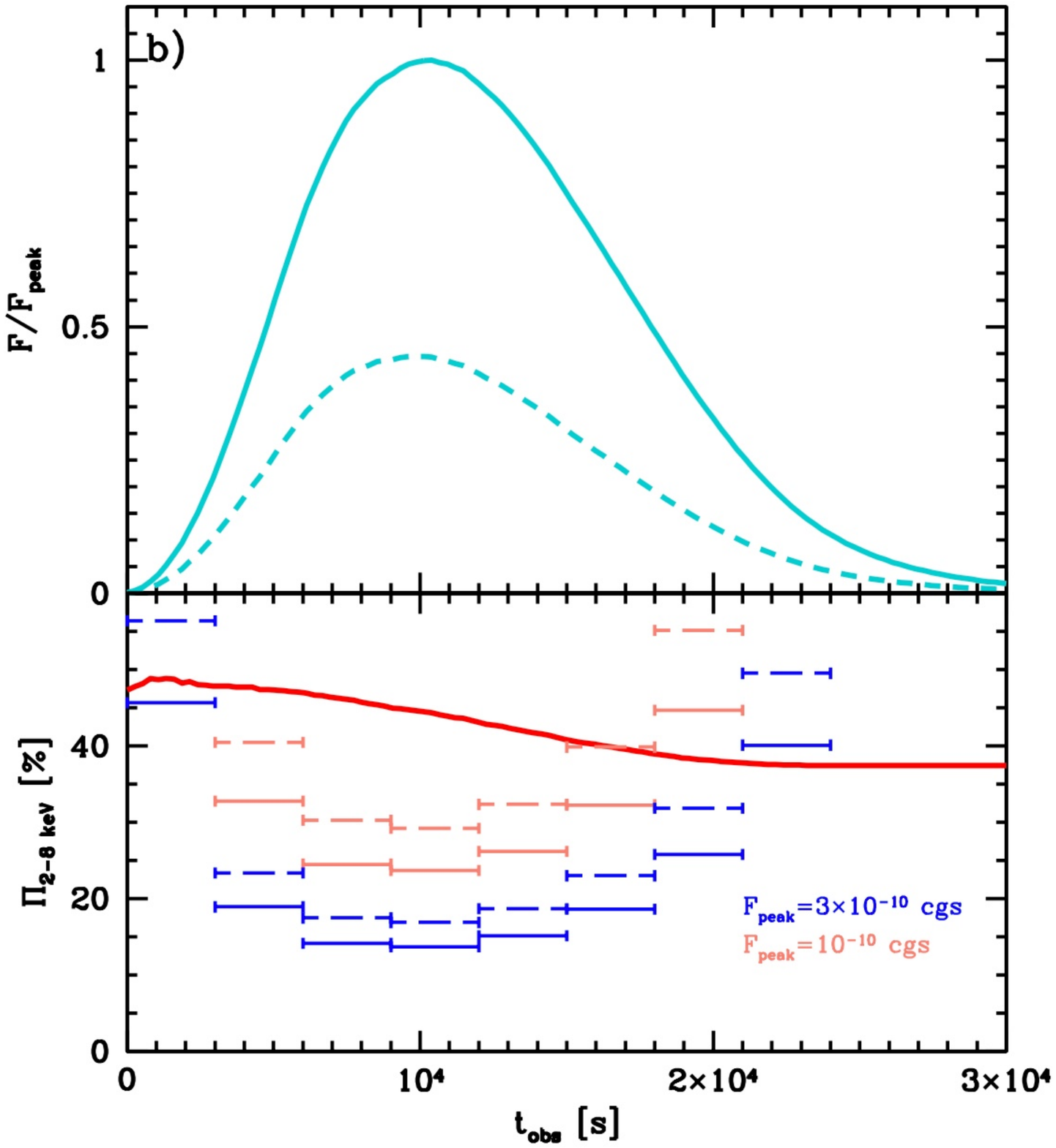}

\caption{(\textbf{a}) Upper panel: normalized light curves at 10 keV (light blue), 1 keV (orange), and in the optical band (green) assuming $m=3$  and an injection (rest frame) timescale $t_{\rm dur}=0.1\times r/c$ (left) and $t_{\rm dur}=r/c$ (right), where $r=10^{15}$ cm is the assumed jet radius. The dashed line shows the polarized flux. Middle panel: the degree of polarization in the three bands. Lower panel: polarization angle in the three bands. All of the quantities are expressed in the observer frame. (\textbf{b}) Lightcurve (upper panel) and degree of polarization (lower panel) in the {\it IXPE} band (2--8 keV) for the case $m=3$ and $t_{\rm dur}=r/c$. In the lower panel we show the minimum detectable polarization (MDP) at 99\% confidence for exposures of 1 ksec, for two peak fluxes and spectral photon index (1.5, solid; 3, dashed). From \cite{tavecchio20}. Reproduced with kind permission from Oxford University Press and the Royal Astronomical Society.}
\label{fig:ixpe}
\end{figure}

Figure~\ref {fig:ixpe} (left panel) reports the resulting time-dependent flux (upper panel), degree of polarization (middle), and angle of polarization (lower), while assuming that the self-generated fields decay with distance from the shock front, being located at $z_{\rm sh})$ as a power law $B\propto (z/z_{\rm sh})^{-m}$, with $m=3$ and assuming two different injection durations , i.e., $t_{\rm dur}=0.1\times r/c$ (left) and $t_{\rm dur}=r/c$ (right), where $r=10^{15}$ cm is the assumed jet radius. The curves are shown for the (observed) energies that correspond to hard X-rays, soft X-rays, and the optical~band.

As mentioned above, the hard X-ray emission, which is produced by energetic, rapidly cooling electrons, is concentrated in a thin layer after the shock, where the magnetic field is largely dominated by the self-generated field. At these energies, the degree of polarization quickly stabilizes around $\Pi\simeq 40$\%.
In the soft X-ray band, on the other hand, the cooling length of the electrons, $\lambda _{\rm cool}$, is longer than the distance over which the self-generated field decays significantly, $\lambda _{\rm decay}$, i.e., $\lambda _{\rm cool} \gtrsim \lambda _{\rm decay}$. Therefore, while in the initial phase, the degree of polarization is similar to that at higher energies, it rapidly decreases, reaching a stationary state at $\Pi\simeq 10\%$. Note that the angle of polarization $\chi$ does not change, since the observed polarization, although diluted, is dominated by the strong self-generated field (parallel to the shock front). At optical frequencies, the situation is different. Indeed, the electrons emitting at these frequencies are characterized by a cooling length comparable with the full extension of the downstream region and much larger than the decay length of the self-generated field, $\lambda _{\rm cool} \gg \lambda _{\rm decay}$. Therefore, after an initial phase that is characterized by an evolution similar to that of the X-ray, the degree of polarization monotonically decreases and reaches $\Pi=0$ (when the emission produced in regions with magnetic field orientation differing by $\pi/2$ exactly balances) and then it increases again. The time at which $\Pi=0$ is marked by a flip of the angle of polarization by $\Delta \chi=\pi/2$, flagging the fact that the emission is now dominated by region in which the magnetic field is oriented along the jet axis (i.e., the original upstream magnetic field).

The qualitative behavior that is discussed above is quite robust and it does not depend on the details of the set-up (e.g., profile of the field, duration of the event). Therefore a natural consequence of this scenario is a relatively large polarization in the (medium-hard) X-rays, a prediction soon testable by {\it IXPE}. Figure~\ref{fig:ixpe} (right panel) shows a simulation of the expected performances of {\it IXPE} assuming as a target Mkn 421. The simulation nicely shows that  {\it IXPE} can easily follow the evolution of the polarization during a flare. These observations, complemented by polarimetric monitoring in other bands (in particular optical), will soon allow us to test the model.

The simplified scenario that is described above has important caveats to consider. Among the most important points worth mentioning is that it assumes a perfect planar geometry, a likely good approximation for internal shocks that are produced by variations of speed within the flow but not suitable to describe oblique shocks induced by recollimation (e.g., \cite{komissarov97,nalewajko12,bodo18}). Moreover, in the downstream region, we do not consider turbulence. Therefore, electrons are simply advected by the flow and do not diffuse. Similarly, the self-generated field is not disturbed by possible turbulence that could (at least partially) perturb its ideal geometry, likely reducing the expected degree of polarization. Finally, while the study described here provides a first glimpse of the expected behavior, a more complete exploration of the parameter space is necessary to compare the theoretical expectations with observations.
\\

An important role for turbulence was assumed in \cite{marscher14}. The scenario at the base of the model (see {Figure}~\ref {fig:marscher}) assumes that highly-turbulent plasma flowing in the jet crosses a standing conical shock (that is easily produced through the jet recollimation by an external medium, e.g., \cite{daly88,bodo18}). At the shock, particles are accelerated with an efficiency that depends on the angle of the local magnetic field relative to the shock front. In this model, the variability of flux and polarization observed in blazars is directly determined by the turbulent nature of the flow. Quite interestingly, despite the fact that the emission from turbulent cells is uncorrelated, the model reproduces coherent rotations of the polarization angle (also see \cite{kiehlmann17}). A prediction of the model is also that the time-averaged degree of polarization increases with energies, since the volume filling factor of electrons at the highest energies is very small and, thus, only a few turbulent cells contribute at high frequency. The model was particularly tailored for powerful blazars, for which the X-ray emission is produced by inverse Compton scattering by electrons of relatively low energy. 


As noted above, quite generally, in any scenario where turbulence plays a major role in determining the properties of the emission (in particular, of the polarization), one expects that the number of turbulent cells contributing to the emission depends on the  frequency. In particular, for HSP, we can expect that the optical band receives the contribution of many cells, while only a few cells have the right, somewhat extreme, parameters (magnetic field, maximum energy of the electrons) that are suitable for providing a substantial contribution in the X-ray band (e.g., \cite{peirson18}). In this scheme, one can infer the degree of polarization (that is expected to scale as $\Pi_{\nu }\propto 1/\sqrt{N_{\nu}}$, where $N_{\nu}$ is the number of cells contributing at a given frequency) in the X-ray band once that in the optical is known and a relation between $\nu$ and $N_{\nu}$ is assumed. For a specific set of assumptions, \cite{liodakis19} found that, for typical HSP (Mkn 421, 1ES1959+650), the predicted degree of polarization is not much larger than that corresponding to the optical emission, i.e., $\Pi_X \lesssim 10\%$. Clearly, observations by {\it IXPE}  will soon confirm or reject this prediction.

\end{paracol}
\nointerlineskip
\begin{paracol}{2}
\switchcolumn

\begin{figure}[H]

    \includegraphics[width=10. cm]{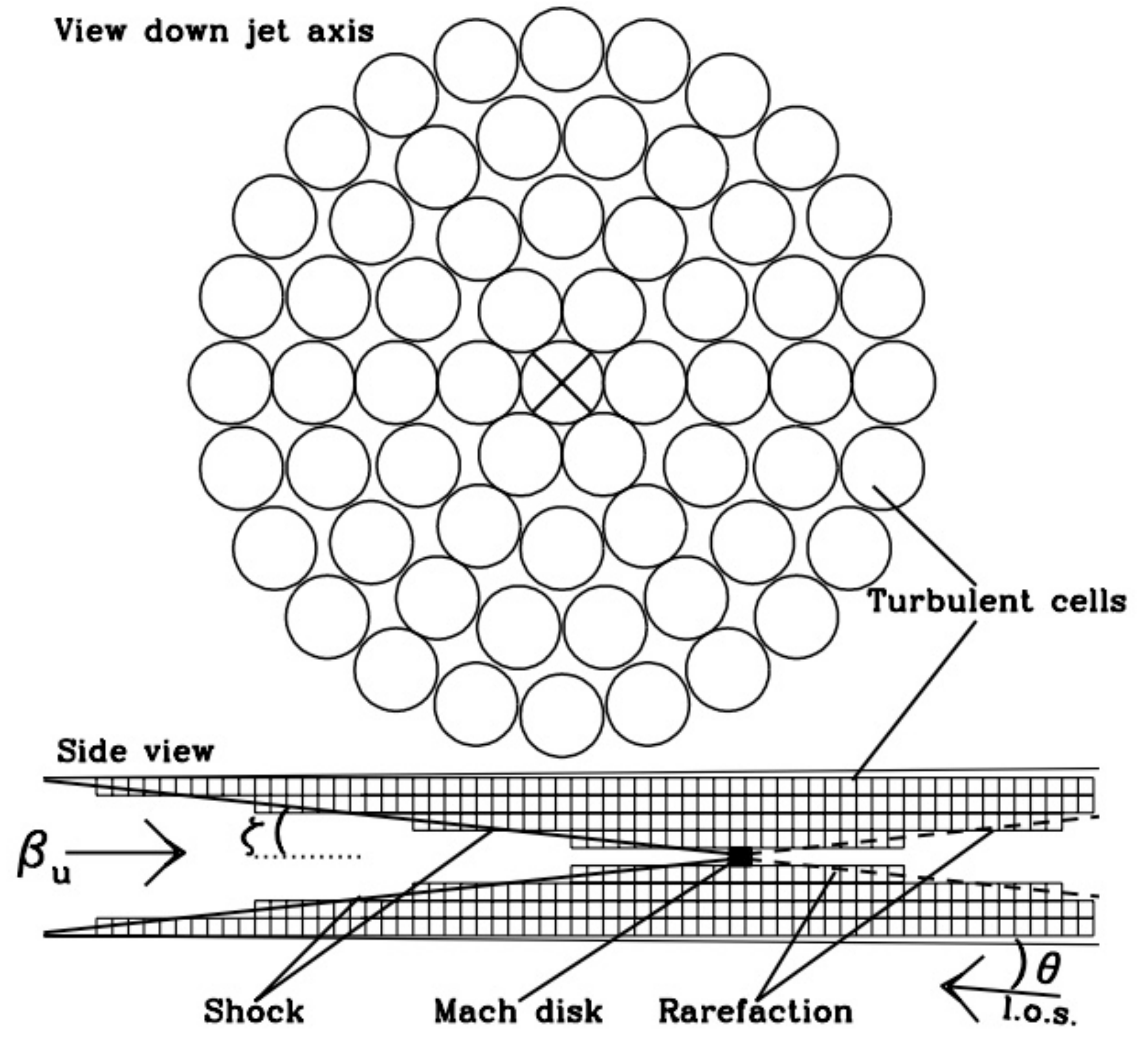}

\caption{A cartoon of the model considered in \cite{marscher14}. Plasma flowing in the jet encounters an oblique shock, where particle acceleration occurs through DSA. In the downstream region the flow is assumed to be highly turbulent and it is modeled by assuming cells in which the physical parameters (in particular, the magnetic field orientation and maximum energy of the electrons) can assume different values. From \cite{marscher14}. Reproduced by permission of the AAS.}  
\label{fig:marscher}
\end{figure}   

\subsection{Magnetic Reconnection in Unstable Jets}

The possibility that particles are boosted to relativistic energies in current sheets, where magnetic energy is released through the reconnection of field lines with opposite polarity, is currently supported by extensive PIC simulations. For high magnetization ($\sigma\gg 1$), the system is indeed able to produce particles with energy distributions showing a prominent power law tail extending to relativistic energies, whose slope is linked to the plasma magnetization (the spectra are harder for larger magnetization, \cite{werner17}). 
Recent studies \cite{petropolou16, christie19, petropolou19, christie20} consolidated the application to blazars of the scenario based on particle acceleration in current sheets. In this framework, one can reproduce the main observational properties of the emission, including the ultra-fast events that were recorded in gamma rays \cite{giannios13,zhdankin20}. Geometries favorable for efficient reconnection are likely to develop under the effect of macroscopic fluid instabilities of the flow, i.e., current driven (kink) instability or Kelvin--Helmholtz instability at the jet/environment interface \cite{hardee13,bromberg19,davelaar20,bodo21,zhang17}

The number of studies of the polarimetric properties of the synchrotron radiation produced by electrons accelerated through magnetic reconnection is still limited. From the point of view of polarimetric measurements, it is conceptually important to distinguish between the effects that are related to the small-scale structure of the magnetic field associated with a single current sheet from the imprints of large-scale fields shaped by the macroscopic dynamics of the system. For instance, recent specific PIC studies \cite{zhang18,zhang20,hosking20} focus on the polarimetric properties of the synchrotron radiation produced by particles diffusing in the complex field that is associated with a single current sheet. The derived synthetic lightcurves reveal large (exceeding $50\%$) and variable polarization fractions and ample variations of the angle of polarization (up to $\Delta \chi=180^{\circ}$). However, the largest timescale of these variations, $\Delta t$ is of the order of the light crossing time of the largest plasmoids that formed in the system, $\Delta t\lesssim 0.1 L/c$, where the size of the system $L$ is of the order of the jet radius $r_j$. For typical physical parameters, this timescale is of the order of $\Delta t\approx 10^3$ s, barely accessible by observations in the X-ray band. This implies that, even if the degree of polarization assumes large values during the evolution, the plane of polarization of the radiation recorded by our instruments will inevitably rotate by a large angle during a single exposure, resulting in an averaged smaller degree of polarization. Moreover, one should consider that several current sheets with different orientations could be active at the same time, further diluting the polarization of the emerging radiation. Therefore, the comparison of the prediction of these small-scale models with the results of actual polarimetric observations necessarily involves the time average of the polarimetric quantities. 
Studies that are based on the fluid MHD simulations focus on the large scale description of the system (e.g., \cite{zhang17,zhang18,bodo21}). This approach is unable to resolve the structure of the current sheets, but, on the other hand, can trace the long-term development of the large-scale topology of field lines in the the entire system. High-energy particles that are injected at reconnection sites can therefore be followed and their emission can be properly modeled when considering the local (but still at the fluid scale) magnetic fields. Under these conditions, the variations of the properties of the emitted radiation occur on relatively long timescales, of the order of the instability growth time and/or light-crossing time of the jet, leading to changes of the polarization angle slow enough to be observationally resolved and, therefore, allowing for tracking the evolution of the polarization.

The investigation of the polarization signatures expected for a jet susceptible to the current driven kink instability has been described in e.g., \cite{zhang17,bodo21}. In the following, we briefly describe the results that were reported in \cite{bodo21}, which explicitly discusses the expected polarimetric signatures for the X-ray band.

The system is idealized as a column of plasma with a magnetic field with both poloidal and toroidal components. It is well known that, when the ratio between the toroidal and the poloidal components exceeds a given value (that is dictated by the detailed structure of the magnetic field), the column is subject to current-driven instabilities, among which the most important is the so-called kink mode (e.g., \cite{begelman98}).

The onset of the instability produces a helical deformation of the jet that, in turn, leads to the formation of a complex system of current sheets (see Figure~\ref{fig:cut}), where magnetic energy is rapidly dissipated through reconnection.  As demonstrated by PIC simulations, a sizable fraction of the released energy is channeled into high-energy particles following nearly power law energy distributions. The simulation assumes that high-energy electrons are injected at current sheets and it follows these particles while they are advected by the flow (diffusion, possibly induced by turbulence, is negligible at least in the early phases) allowing to calculate the properties of the emitted synchrotron radiation. While cooling is not explicitly taken into account, the emission at different energies is estimated when comparing the time of injection of the electrons and their cooling time. In this way, one can consider the emission of both freshly accelerated electrons (assumed to emit in the X-rays) and long-lived ones, emitting at low frequencies, as shown by Figure~\ref{fig:cut}. The particles are confined in the current sheet (at least during the early phases of the evolution, when the luminosity of the produced radiation is maximal) and do not diffuse outside.

\begin{figure}[H]

    \includegraphics[width=13 cm]{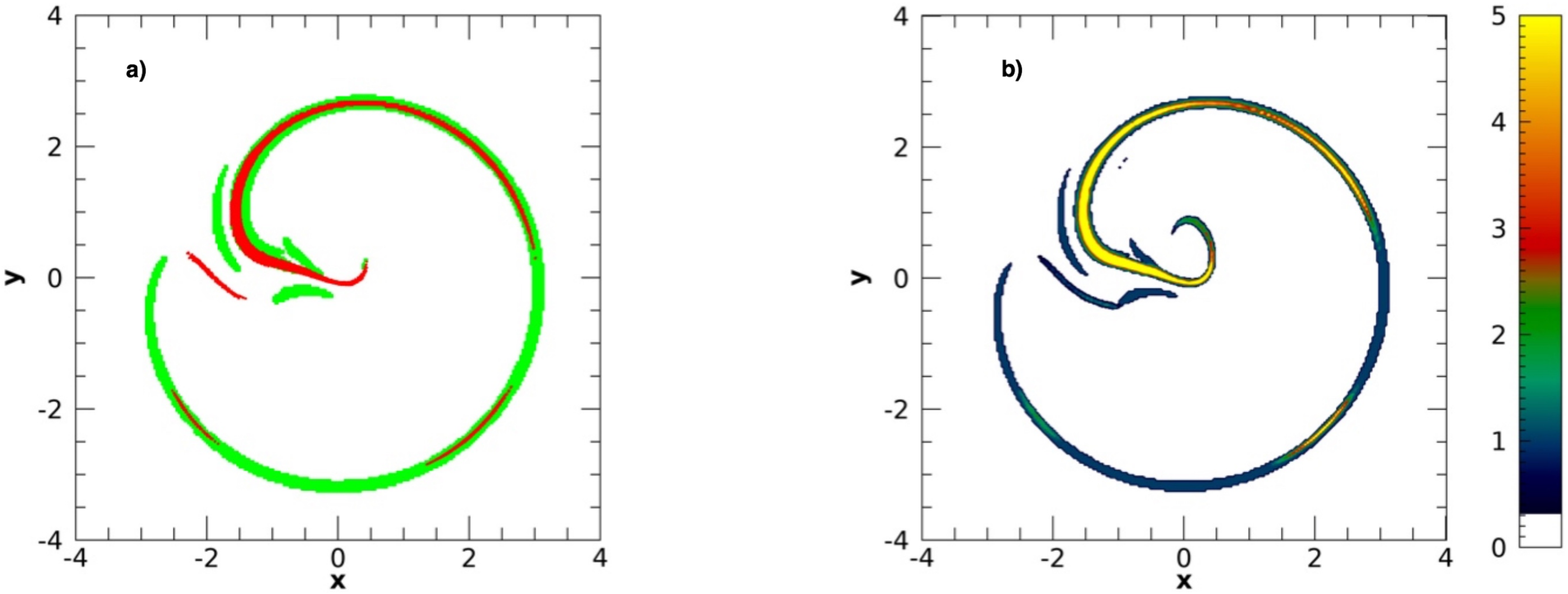}

\caption{(\textbf{a}) The positions of the particles in a transverse section of the jet, in green the particles that have an age less than $r_j/c$ (still not cooled and, hence, assumed to emit high-energy radiation) and in red the particles that have an age between $3r_j/c$ and $10 r_j/c$ (cooled, emitting in the optical). (\textbf{b}) Plot of the the density of particles in the same section of the jet. The panels refer to the time at which the dissipation is maximal. From \cite{bodo21}. Reproduced with kind permission from Oxford University Press and the Royal Astronomical Society.}  
\label{fig:cut}
\end{figure}

Figure~\ref{fig:reconn} shows the resulting properties of the emission. After a short transient phase, the degree of polarization settles around $20\%$. This relatively low value is determined by the contribution of several active current sheets with different orientation that determine an effective dilution of the total polarization. Remarkably, X-rays and optical frequencies are characterized by quite similar values and evolution of $\Pi$. This is clearly related to the fact that, as discussed above, particles are confined in the current sheet and therefore probe very similar magnetic field structures. If, as suggested by recent works, (e.g., \cite{davelaar20}), a relatively high-level of turbulence develops at late times, the corresponding enhanced diffusion of particles could induce appreciable differences between the emission of freshly injected particles (emitting at X-rays) and that associated with old electrons. 

A quite important feature is that the polarization angle displays important changes, with multiple rotations by $\Delta \chi \approx 90^{\circ}$. As stressed above, these changes are slow enough to be followed with current instruments, thus making it possible to measure the degree of polarization and its evolution during a flare.

While these simulations already offer an interesting view of the expected polarimetric signatures that are associated with magnetic reconnection induced by kink instability, a more robust investigation should include a better treatment of the particle injection and cooling and transport.  

\begin{figure}[H]
    \includegraphics[width=9.5 cm]{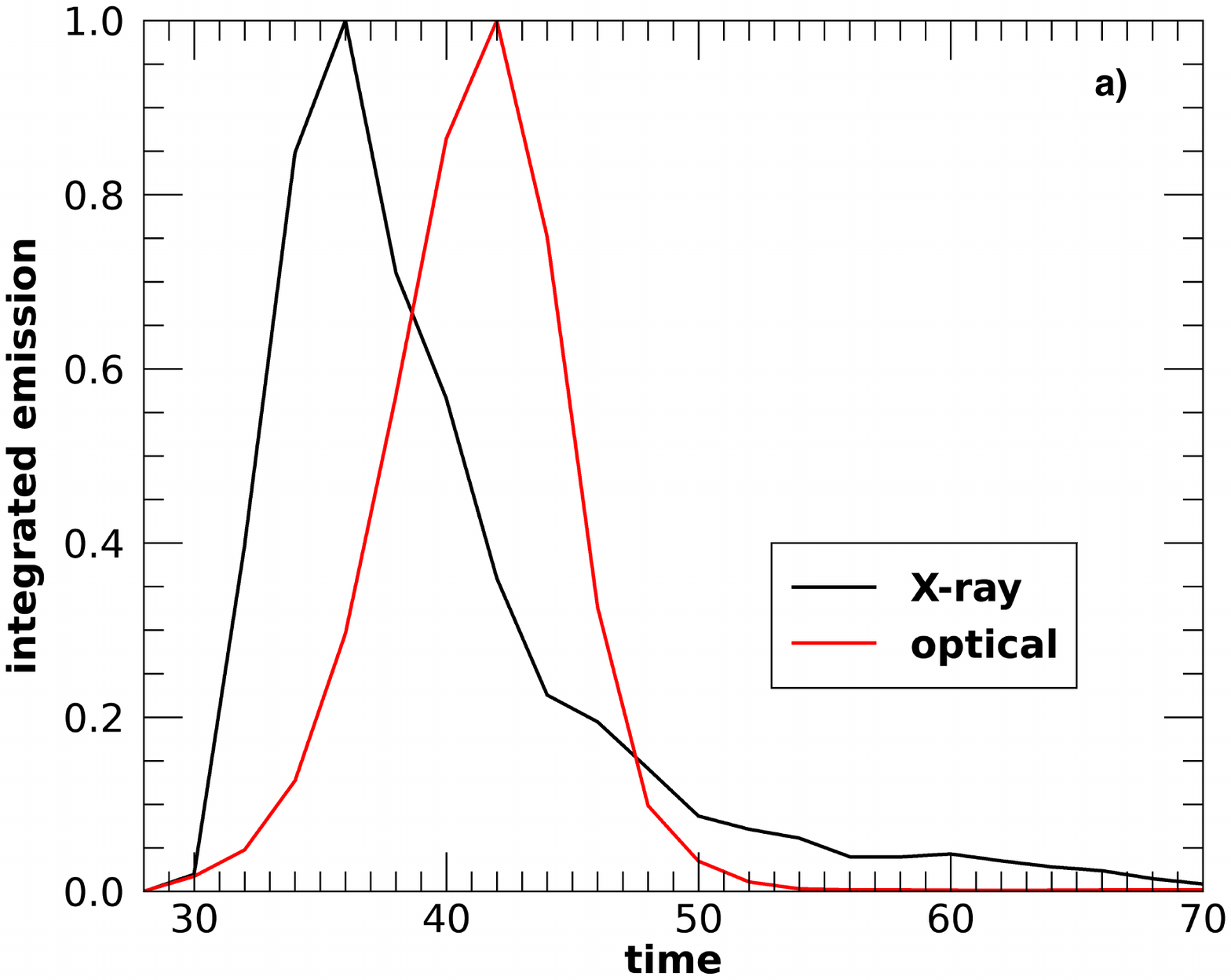}
    \hspace{-3truecm}
    \includegraphics[width=9.5 cm]{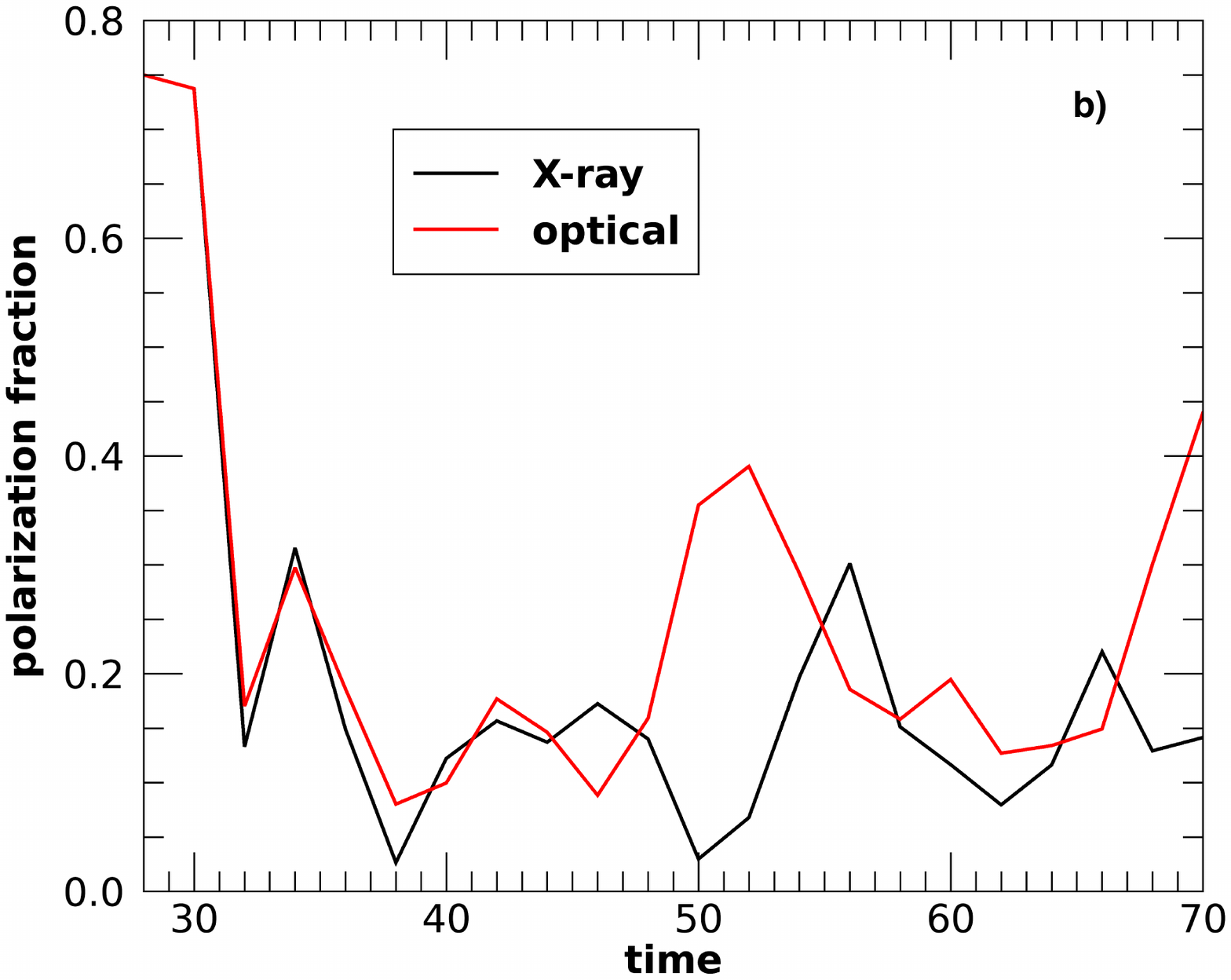}
        \hspace{-2.2 truecm}
    \includegraphics[width=9.5 cm]{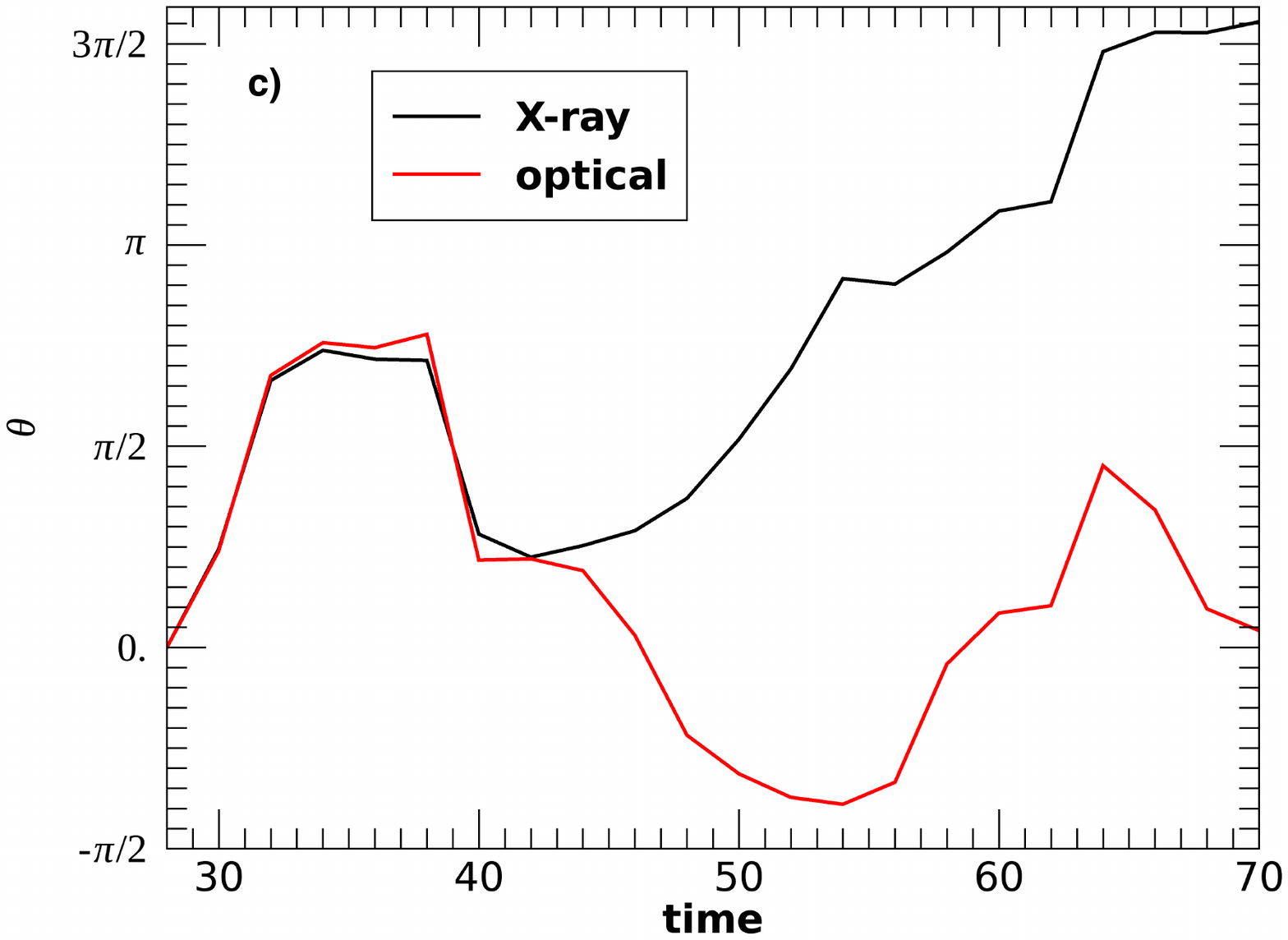}
\caption{(\textbf{a}) Plot of the total emission in the X-ray (black) and optical (red) bands as a function of time for the simulation presented in \cite{bodo21}. Each curve is normalized to its peak value. (\textbf{b}) Polarization fraction in  the X-ray (black) and optical (red) bands as a function of time. (\textbf{c}) Evolution of the polarization angle of the integrated emission as a function of time (colors as above). From \cite{bodo21}. Reproduced with kind permission from Oxford University Press and the Royal Astronomical Society.\label{fig:reconn}}
\end{figure}

\section{Outlook}

The polarization measurements of the X-ray synchrotron emission from blazars of the HSP class promise to shed light on the basic mechanism(s) responsible for energy dissipation and particle acceleration in jets. In this review, we focus on the anticipated polarimetric properties for two potential mechanisms, i.e., diffusive shock acceleration and magnetic reconnection.

As clearly shown in Figures \ref{fig:ixpe} and \ref{fig:reconn}, the two specific scenarios that are discussed above, i.e.,  DSA with self-produced magnetic fields and magnetic reconnection that is induced by kink instability, lead to quite different predictions for the polarimetric properties in the X-ray band (a summary is reported in Table \ref{table:table}): 

\begin{enumerate}
\item[(1)]	For acceleration by shocks that are characterized by a magnetic field with a strong self-produced component, one anticipates a quite large (around $40\%$ for the parameters assumed in \cite{tavecchio18,tavecchio20}) and stable degree of polarization of the X-ray emission. Moreover, since the emission occurs in a region that is characterized by a well defined orientation of the magnetic field, the angle of polarization is not expected to display large changes during the evolution of a flare. However, at optical frequencies, one predicts a more complex behavior. It is important to note that, while, for the X-ray band, it is likely that the emission is dominated by a single component (especially during flares), in the case of the optical emission it is quite likely that several components (or large portion of the jet) can contribute to the observed flux. This likely implies a dilution of the measured polarization in the optical, in agreement with the usually low (around 10$\%$) average degree of polarization measured in BL Lac objects. As discussed above, this simple model neglects the possibly important role of turbulence in the downstream flow. A high level of turbulence can destroy the order of the self-generated field, greatly affecting the resulting polarization.
\item[(2)]	In the case of flares that are powered by the dissipation of magnetic energy in current sheets produced during the evolution of instabilities one expects a relatively small polarization (around 20$\%$) at all bands, as a result of the simultaneous contribution of several active current sheets with different orientation. Moreover, the evolution of the instability results in 
significant variations of the angle of polarization over timescales that {\it IXPE} can easily resolve. The confinement of particles within the current sheets mainly determines the similar polarization in the optical and the X-ray band. However, the development of turbulence, which results in the effective energy-dependent diffusion of particles, could have an important role in shaping the polarimetric properties. Further studies are required to clarify the situation. 

\end{enumerate}


\end{paracol}
\nointerlineskip
\begin{specialtable}[H] 
\widetable
\caption{Summary of predicted polarization features for some of the scenarios discussed in the text.} \label{table:table}

\setlength{\cellWidtha}{\columnwidth/3-2\tabcolsep+0.0in}
\setlength{\cellWidthb}{\columnwidth/3-2\tabcolsep+0.0in}
\setlength{\cellWidthc}{\columnwidth/3-2\tabcolsep-0.0in}
\scalebox{1}[1]{\begin{tabularx}{\columnwidth}{>{\PreserveBackslash\centering}p{\cellWidtha}>{\PreserveBackslash\centering}p{\cellWidthb}>{\PreserveBackslash\centering}p{\cellWidthc}}
\toprule
 \T & \textbf{Optical} & \textbf{Medium-Hard X-Rays} \B  \\
\midrule
Shock (turbulent) \T  & $\Pi\lesssim 15\%$, variable; &  $\Pi\lesssim 30\%$, highly variable    \\
& $\chi$ variable, smooth rotations possible & highly and rapidly variable  \\
&  &   \\
Shock (self-produced field) & $\Pi \lesssim 20\%$, slowly variable, & $\Pi\ \gtrsim 40\%$ substantially constant,\\
&  flips by $\Delta \chi=90$ deg & constant $\chi=0$\B  \\
&  &   \\
Reconnection (kink-induced)& $\Pi\lesssim$ 20--30\%, moderately variable & same as optical \B   \\
&  smooth rotations,  $\Delta \chi \gtrsim 90$ deg  & as optical\B  \\
\bottomrule
\end{tabularx}}
\end{specialtable}
\begin{paracol}{2}
\switchcolumn

Despite the high degree of simplification, the two scenarios that are described above represent solid theoretical benchmarks that can be easily contrasted with future data. In particular,  {\it IXPE} (planned to be launched at the end of 2021) will allow us to study in great detail the polarimetric properties of HSP, providing us an unprecedented test bed for our models.


\vspace{6pt} 

\funding{This research received no external funding.}

\acknowledgments{I would like to thank Lorenzo Sironi, Gianluigi Bodo, Marco Landoni and Paolo Coppi for years of fruitful collaboration in studying X-ray polarization from jets and for fruitful discussions.}
\conflictsofinterest{The author declares no conflict of interest.}

\end{paracol}
\reftitle{References}





\end{document}